\documentclass[reprint,superscriptaddress,amsmath,amssymb,amsfonts,aps,prapplied,floatfix]{revtex4-1}

\usepackage[T1]{fontenc}
\usepackage[utf8]{inputenc}
\usepackage{xcolor}
\usepackage{graphicx}
\usepackage{float}
\usepackage{ulem}
\usepackage{url}
\usepackage[colorlinks=true, linkcolor=blue, citecolor=blue, urlcolor=blue, bookmarks=true, breaklinks=true]{hyperref}
\usepackage{bookmark}
\usepackage{siunitx}
\usepackage{tikz}
\usetikzlibrary{external}
\usepackage{pgfplots}
\pgfplotsset{compat=1.15}
\usepgfplotslibrary{external}
\usepackage{textcomp}

\DeclareSIUnit[]{\wire}{wire}

\begin{document}

\title{
Piezoresistance characterization of silicon nanowires in uniaxial and isostatic pressure variation}

\author{Elham Fakhri}
\affiliation{Department of Engineering, Reykjavik University, Menntavegur 1, IS-102 Reykjavik, Iceland}

\author{Rodica Plugaru}
\affiliation{National Institute for Research and Development in Microtechnologies- IMT Bucharest,
077190 Voluntari, Romania}

\author{Muhammad Taha Sultan} 
\affiliation{Department of Engineering, Reykjavik University, Menntavegur 1, IS-102 Reykjavik, Iceland}
\affiliation{Science Institute, University of Iceland, Dunhaga 3, 107 Reykjavik, Iceland}

\author{Thorsteinn Hanning Kristinsson}
\affiliation{Department of Engineering, Reykjavik University, Menntavegur 1, IS-102 Reykjavik, Iceland}

\author{Hákon Örn Árnason}
\affiliation{Department of Engineering, Reykjavik University, Menntavegur 1, IS-102 Reykjavik, Iceland}

\author{Neculai Plugaru}
\affiliation{National Institute for Research and Development in Microtechnologies- IMT Bucharest,
077190 Voluntari, Romania}

\author{Andrei Manolescu}
\affiliation{Department of Engineering, Reykjavik University, Menntavegur 1, IS-102 Reykjavik, Iceland}

\author{Snorri Ingvarsson}
\affiliation{Science Institute, University of Iceland, Dunhaga 3, 107 Reykjavik, Iceland}

\author{Halldor Gudfinnur Svavarsson}
\affiliation{Department of Engineering, Reykjavik University, Menntavegur 1, IS-102 Reykjavik, Iceland}

\begin{abstract}
Silicon nanowires (SiNWs) are known to exhibit large piezoresistance (PZR) effect, making it suitable for various sensing applications. Here, we report the results of a PZR investigation on randomly distributed and 
interconnected vertical silicon nanowire arrays as a pressure sensor. The samples were produced from 
p-type (100) Si wafers using a silver catalysed top-down etching process. The piezoresistance response of these SiNW 
arrays was analysed by measuring their I-V characteristics under applied uniaxial as well as isostatic pressure. The 
interconnected SiNWs exhibit increased mechanical stability in comparison with separated or periodic nanowires. The 
repeatability of the fabrication process and statistical distribution of measurements were also tested on several 
samples from different batches. A sensing resolution down to roughly \SI{1}{\milli\bar} pressure was observed 
with uniaxial force application, and more than two orders of magnitude resistance variation was determined for 
isostatic pressure below atmospheric pressure.
\end{abstract}


\maketitle

\section{Introduction}

Low dimensional structures may possess unique mechanical, electrical, optical, and thermoelectric properties. 
Particularly, silicon nanowires (SiNWs), have demonstrated properties suitable for various advanced applications, 
\cite{Peng2013,Heris2020,Heris2022}, including also low-cost thermoelectric devices and chemo-biological 
sensors with ultrahigh sensitivity \cite{Zhou2003,Peng2009}. The integration of SiNWs in electronic devices is favored 
by their compatibility with the well established Si-SiO\textsubscript{2} electronic industrial technology. Bulk silicon 
has been known for a while to exhibit high piezo resistance (PZR) effect \cite{Smith1954}. In bulk semiconductors the 
PZR-effect takes place, in principle, due to a change in the electronic structure and modification of the 
charge-carriers effective masses. This phenomenon has found practical applications in many Si-based devices, such as 
pressure transducers \cite{Tufte1962}, cantilevers for atomic force microscopy \cite{Tortonese1993}, accelerometers 
\cite{Ning1995}, biosensors \cite{Wee2005}, and multi-axis force sensing tools \cite{Tiwari2021}. 

Recently, nanowires have been shown to possess the ability to significantly increase the PZR response. A giant PZR was observed in p-doped SiNWs with diameters of \SIrange{50}{350}{\nano\meter} and length of microns, initially under tensile uniaxial stress \cite{He2006}. However, the PZR effect in n-doped nanowires was found comparable to that in the bulk counterpart, both for tensile and compressive uniaxial stress  \cite{Gao2017}.

On the theoretical side, the origin of the PZR effect in SiNWs has long been under debate and most frequently it is 
referred to as anomalous PZR \cite{Zhang2012}. It has been related to quantum confinement effects \cite{Cheng2018}, 
surface charge effects \cite{Nguyen2021,Kim2020,Ghosh2021}, strain-induced bandgap shift \cite{shiri2008}, or changes of 
the charge carriers effective masses \cite{Zhang2016}. A complex model incorporating these mechanisms has been proposed 
in order to analytically quantify the PZR effect in silicon \cite{Rowe2014}.

A survey of several PZR sensors based on SiNWs, e.g. cantilever \cite{Toriyama2002}, opto-mechanical sensor 
\cite{Toriyama2003}, flexible pressure sensor \cite{Kim2020}, or breath detector \cite{Ghosh2021}, shows that different 
methods have been used for fabricating the SiNWs, such as vapor-liquid-solid (VLS), laser ablation, and metal assisted 
catalysed etching (MACE) \cite{Schmidt2009}. Among these methods, MACE is the simplest and most versatile one. It relies 
on catalyzed etching with assistance of a perforated metal template film (typically gold or silver) 
\cite{svavarsson2016} or randomly distributed metallic nanoparticles (typically gold or silver) 
\cite{Fakhri2021,plugaru2022structure} spread  on the Si-wafer. To date, studies have been focused on uniaxial load and 
have neglected the SiNWs response under isostatic pressure, which creates a load uniformly distributed on the 
sample surface. Here, we report on the PZR effect in SiNWs obtained by MACE, under uniaxial compression load, as well 
as isostatic pressure in a vacuum chamber. We find that the interconnected SiNWs are mechanically stronger and 
functionally more stable comparing with the arrays of separated wires under applied uniaxial pressure. They show 
higher PZR sensitivity under isostatic pressure variation. We also demonstrate a simple, low cost and reproducible 
fabrication method, to fabricate a robust and sensitive pressure sensor.

\section{Materials and Methods}

\subsection{Fabrication of SiNWs} 

Arrays of interconnected SiNWs were fabricated by silver (Ag) MACE in a three step process, from p-type, 
single-side polished \SI{525}{\micro\meter} thick Si wafers, with resistivity, $\rho$, of 
\SIrange{10}{20}{\ohm\centi\meter}. The nanowire patterns were made on areas of about \SI{1}{\centi\meter\squared}, on 
the polished side of the wafers. The sequence of steps used to prepare the SiNW areas is as follows:

\begin{enumerate}
\item Deposition of metal catalyst: Ag nano-particles were deposited on the surface of the Si-wafers by immersing the 
wafers in a solution of \SI{3}{M} HF and \SI{1.5}{M} AgNO\textsubscript{3} for \SI{60}{\second}.  

\item Wire etching: The samples were etched by immersing them in HF:H\textsubscript{2}O\textsubscript{2} (5M:0.4M) solution to obtain vertically aligned SiNWs.

\item Removal of residual Ag nano-particles: Samples were immersed in \si{20}{\%} w/v HNO\textsubscript{3} to remove residual silver particles. More detailed description can be found in Refs.~\cite{Fakhri2021,plugaru2022structure}.
\end{enumerate}

Subsequently, \SI{150}{\nano\meter} thick aluminum electrodes were deposited on the samples by electron beam evaporation (Polyteknik Cryofox Explorer 600 LT). For the uniaxial measurements, the electrodes were deposited on top and backside of the samples, while for the isostatic pressure measurements, the electrodes were made co-planar. 

Four sets of interconnected SiNWs samples, denoted as follows, were prepared by varying the etching time: A 
(\SI{1}{\minute}), B (\SI{3}{\minute}), C (\SI{5}{\minute}), D (\SI{7}{\minute}) were made for uniaxial pressure 
application and sample E (\SI{40}{\minute}) was made for isostatic pressure testing. 

Additionally, periodic SiNWs, sample F, was made as described in the Appendix. 
\begin{table}[H]
\caption{List of samples with different etching time and corresponding length. Samples A,B,C,D,E are random wires and sample F is with periodic wires (shown in the Appendix).}
\label{tb:tab1}
\begin{center}
\begin{tabular}{ c c c }
\hline
\textbf{Sample~~~} & \textbf{Etching time (min)~~}	& \textbf{SiNW length (\textmu m)}	\\
\hline
A & 1 & 0.7		   \\
B & 3 & 1.5		   \\
C & 5 & 2.2        \\
D & 7 & 3          \\
E & 40 & 10    \\
F & 5 & 0.65       \\
\hline
\end{tabular}
\end{center}
\end{table}

\begin{figure}
    \centering
    \includegraphics[width=0.8\linewidth]{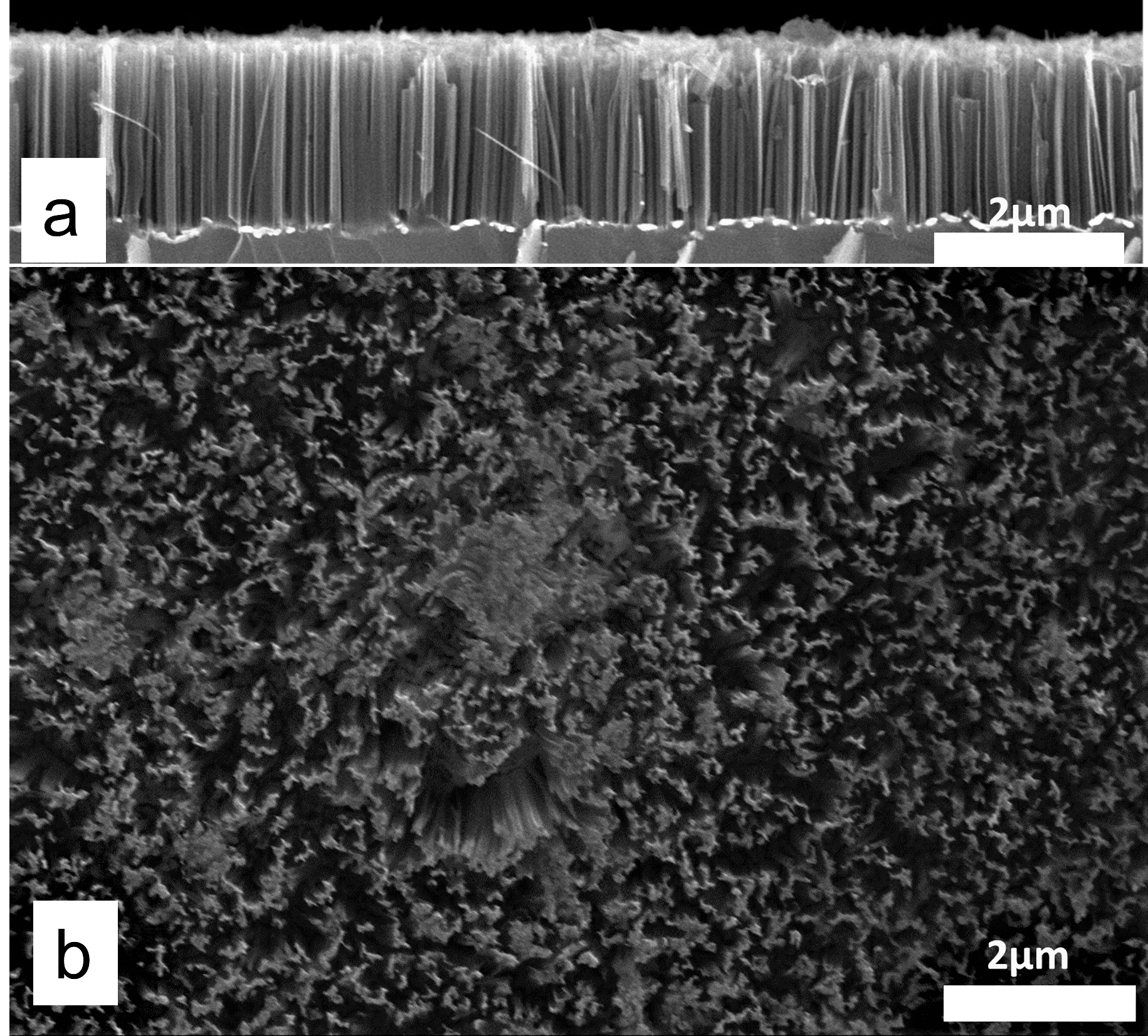}
    \caption{(a) Cross-sectional SEM image of SiNWs, (b) top-view SEM image of SiNWs etched for \SI{7}{\minute}}
    \label{fig:SEM}
\end{figure}

A scanning electron microscope (SEM, Zeiss Supra 35) was used to characterize the SiNWs geometry. Top and cross-sectional SEM images were used to estimate the average diameter of the wires. 
The Gwyddion software for data visualization and analysis was applied to surface SEM images in order to estimate the total top-area of the wires.
\autoref{fig:SEM} shows (a) the cross-section of wire array, and (b) the surface area of SiNWs obtained after \SI{7}{\minute} etching. It is worth mentioning here that the SEM analysis was carried out on the sample prior to removal of Ag-nanoparticles which are visualized as bright spots at the base of the nanowires in the cross-section.
\begin{figure}
  \centering
  \includegraphics[width=0.8\linewidth]{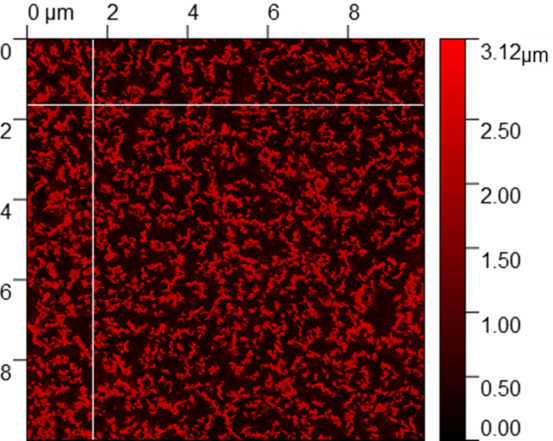}(a)
 \includegraphics[width=0.72\linewidth]{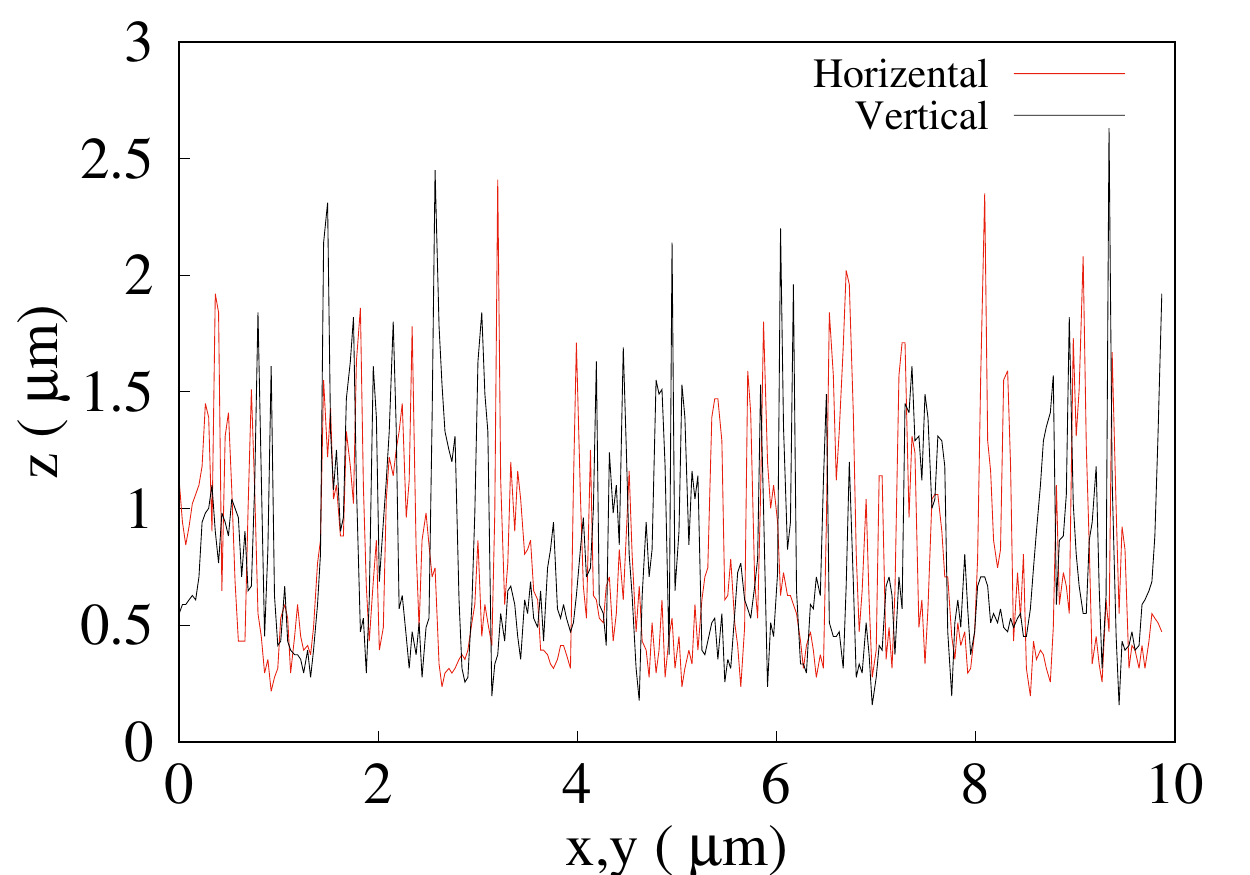}(b)\hspace{17mm}
  \caption{(a) Gwyddion analysis of top-view SEM image in (\SI{100}{\micro\meter\squared}). (b) Two Gwyddion line-scans of the figure in left  (red color - Horizontal line,  black color - Vertical line).}
   \label{fig:Gwyddion}
\end{figure}
As can be seen in the top-view image, the wires are partly interconnected, forming a continuous rigid structure. Also seen from  place to place are free standing nanowires forming bundles. 
Such bundle formation may take place because of capillary forces acting during the drying process following the wet-etching step.
In the cross sectional image one may observe that the length of the wires is relatively homogeneous, around 
\SI{3}{\micro\meter} and their typical diameter is approximately \SI{150}{\nano\meter}. 
According to Peng et al. \cite{Peng2009}, porosity plays an essential role in the PZR response, that increases with increased porosity. 
The porosity is most conveniently controlled by the concentration of Ag deposition solution and etching time. 
In a previous study \cite{Fakhri2021}, it was demonstrated that the SiNWs porosity was highly affected by the concentration of the AgNO\textsubscript{3} during the Ag-deposition. 
A \SI{1.5}{mM} AgNO\textsubscript{3} (as used for samples A-D) provided highly porous SiNWs with maximum photoluminescence spectra intensity. 

From the SEM image analysis, the wires cross-sectional surface coverage was estimated as roughly \si{28}{\%} by using 
the Gwyddion program, \autoref{fig:Gwyddion}. 
By counting the average number of wires on several line-scans, the wire density was estimated to be \SI{1.6 e7}{\milli\meter\squared}, which corresponds to roughly \num{8 e8} wires under the force meter area (\SI{7 x 7}{\milli\meter}).\\

\subsection{Measurement setups} 

\begin{figure}
    \centering
   \includegraphics[width=0.6\linewidth]{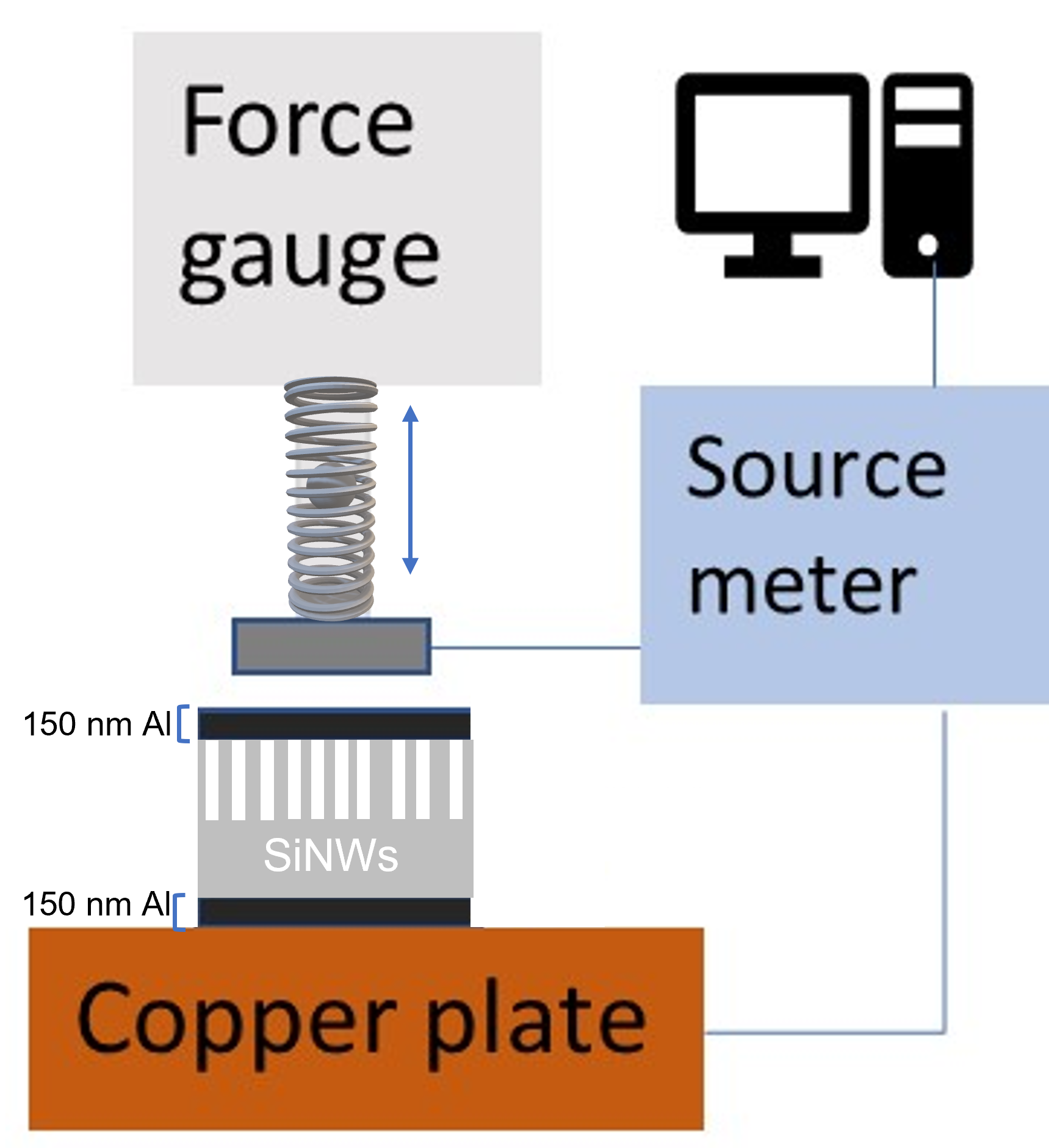}
    \caption{Schematic of the force meter setup for testing PZR characteristics.}
    \label{fig:schem}
\end{figure}

For uniaxial PZR tests, the samples were clamped between a metal pin of a force meter (Mark-10, M5-012), touching the top-side of the samples, and a rigid copper (Cu) plate at the backside as shown in \autoref{fig:schem}. In order to improve the electrical contacts, the samples were glued to the Cu backside plate with silver paste. The force meter was mounted on a vertical rod allowing for a movement on the z-axis (vertical) by a manually adjusted screw. A uniaxial pressure was applied to the samples by pressing the pin of the force meter into the samples' surface with intensity determined by the aforementioned screw. 
The applied force was in the range of \SIrange{100}{900}{\milli\newton} on an area of \SI{7 x 7}{\milli\meter} area (the cross sectional area of the pin). Taking the wire coverage (\SI{28}{\%}) into account, this corresponding to a gauge pressure of \SIrange{7}{66}{\kilo\pascal}, respectively.
\autoref{fig:schem} shows sketch of the experimental setup. 
A Keithley 2400 SourceMeter was used to measure the resistance $R$ through the sample as a function of applied pressure, at a constant bias voltage of \SI{2}{\volt}.

For PZR tests under isostatic pressure variation a vacuum chamber was used, and the air was removed while the resistance was measured at fixed \SI{5}{\volt}.  In this case (for sample E) the contacts were made in co-planar configuration, with separate contacts on each side of the patterned area, such that the nanowires are maximally exposed to air, instead of top-and-bottom contacts as for the uniaxial measurements (performed with samples A-D).
Furthermore, the wire length was increased (to \SI{10}{\micro\meter}) to increase the surface area exposed to the air.
The samples were mounted on a fixed sample station inside the chamber with tungsten needle tip kelvin probes as a connection. 

\section{Results}

\subsection{Electrical response under uniaxial force}

A maximum force of approximately \SI{900}{\milli\newton} was applied to the samples using the setup shown in \autoref{fig:schem}.
The maximum  vertical  force exerted on each wire thus corresponds to \SI{900e-3}{\newton}/\num{8e8} \SI{\approx 1.1e-9}{\newton/\wire} or \SI{6.6e 4}{\pascal} (\SI{0.65}{\bar}). 

\begin{figure}
  \centering
    \includegraphics[width=0.8\linewidth]{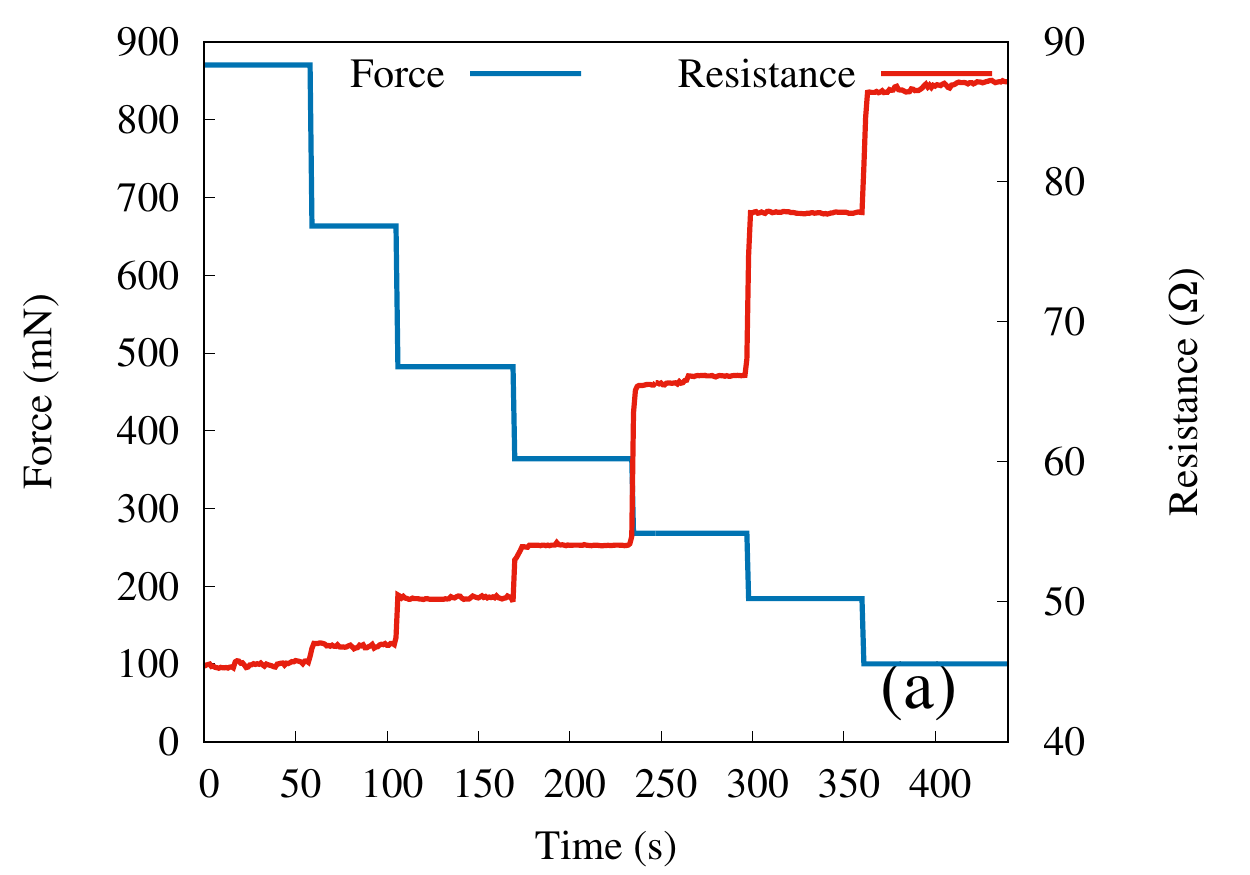}
   \includegraphics[width=0.85\linewidth]{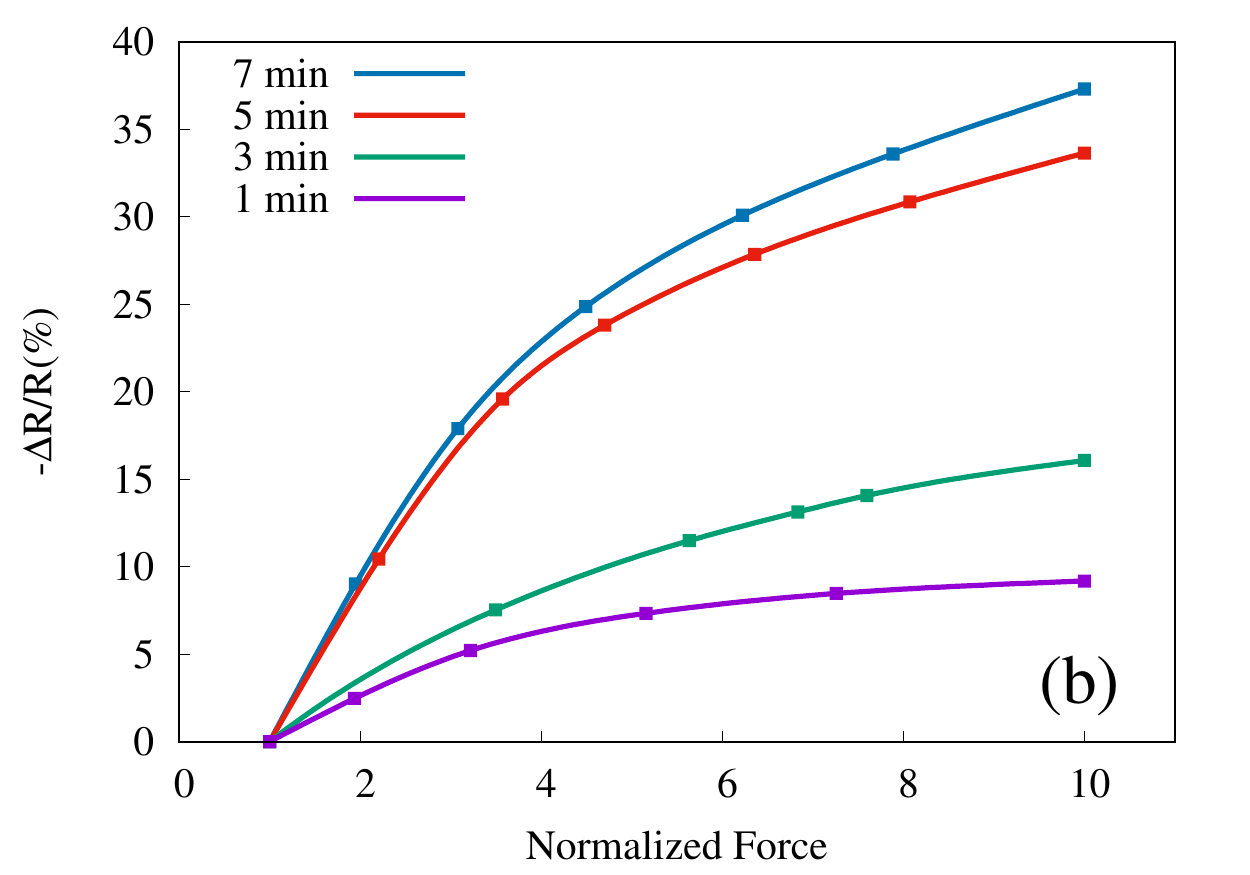} 
   \caption{(a) Instantaneous resistance (right y-axis, red line) in response to applied force (left y-axis, blue line) over time for sample D.  (b) Relative resistance change versus normalized force for SiNWs samples A (purple), B (green), C (red), D (blue).}
   \label{fig:Force}
\end{figure}

The force and the instantaneous resistance measured after applying the force on the sample with nanowires of length \SI{3}{\micro\meter} (etched for \SI{7}{\minute}),  are shown in \autoref{fig:Force}(a), as function of time. The maximum force was applied at the beginning (time zero) and then reduced step-wise while the resistance was being measured. Each force level was kept constant for roughly 50~s to confirm that the resistance value was stable with time. As clearly visible, the resistance changes significantly and inversely with pressure.

We define the relative resistance change vs force variation, for each step $k=1,2,...$ as 
\begin{equation}
   \frac{\Delta R}{R}=\frac{R_k-R_1}{R_1},
   \label{eq:first}
\end{equation}
where $R_k$ is the average resistance obtained for the constant force step $k$.
The measurements were repeated four times on each sample (A, B, C, D), every time reproducing the force steps as well as possible with the adjustable screw.
A second average, this time of the relative resistance in step $k$, defined by \autoref{eq:first}, obtained for the four measurements, vs. the average normalized forces for each step, are presented in \autoref{fig:Force}(b).
Here, by normalized force we mean the ratio of the force in a particular step $k$ to the initial force, $F_k/F_1$.  Hence, the maximum force values for each step-wise measurement are normalized to unity.
Note that the resistance decreases when the force increases, and that in \autoref{fig:Force}(b) we show the negative of $\Delta R / R$.   

Increasing the length of the nanowires (by increasing the etching time) leads to higher relative resistance, although the trend appears to saturate; only a relatively small difference between the \SI{5}{\minute} and \SI{7}{\minute} samples has been obtained.
The highest relative change (\SI{37.3}{\%}) in resistance is observed for the longest etching time (sample D) with regards to uniaxial pressure.
\begin{figure}
    \includegraphics[width=0.8\linewidth]{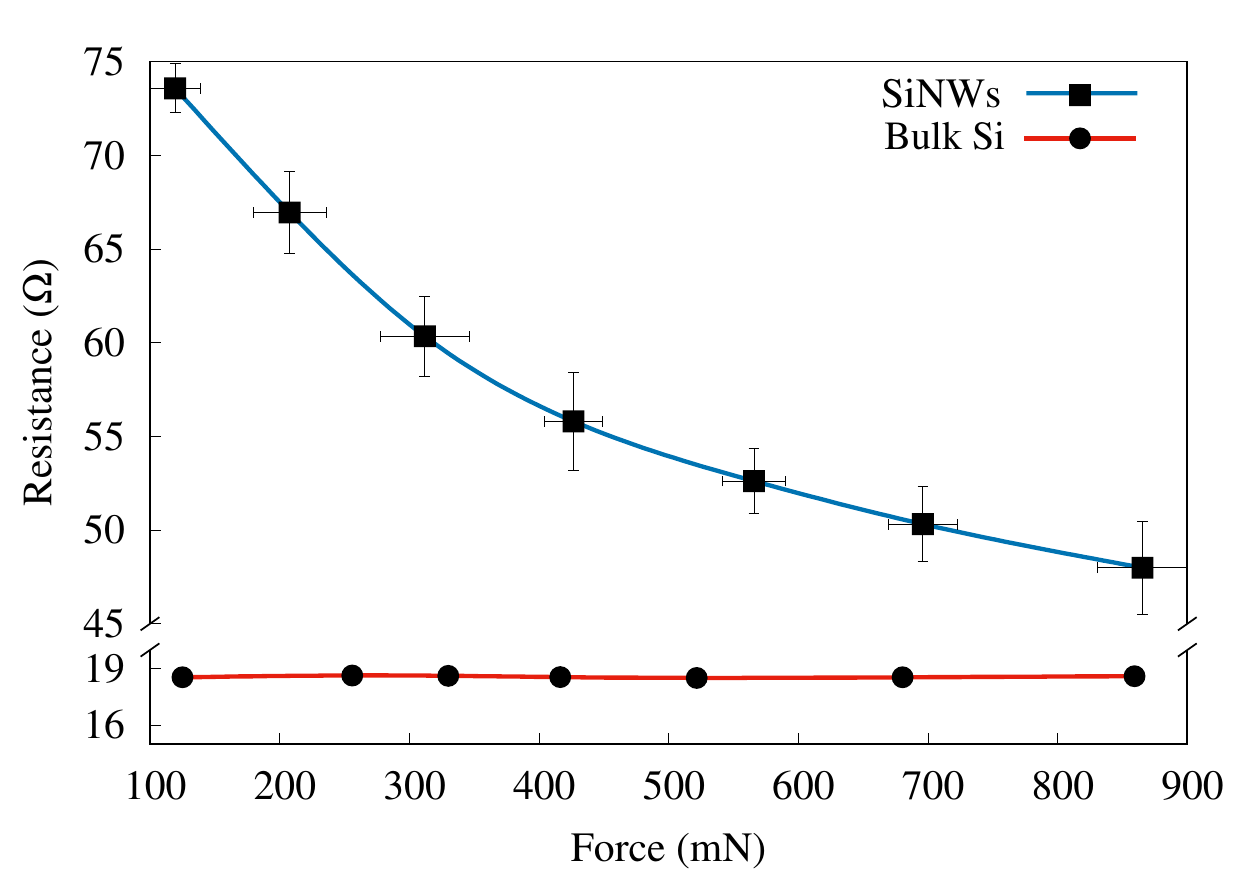}
   \caption{Resistance vs. applied force for the D-like samples (blue line) and bulk-Si (red line). Data for SiNWs are average values for 4 distinct samples, error bars show the range in  which the measurements fell in. 
   }
   \label{fig:RandF}
\end{figure}
In \autoref{fig:RandF} the resistance versus applied force is shown again for sample D (blue line), now compared with the case of the bulk Si (red line).
For this measurement we replicated the sample D four times, in the same conditions of 7 min etching time, and performed the measurements on each sample replica.
The data displayed are average values measured on these samples (D-like), shown with the error bars.  A striking difference between these two configurations (SiNWs vs. bulk Si) is observed. No measurable change in resistance is seen 
for the bulk Si, in a very sharp contrast to the SiNWs sample.

\begin{figure}
    \centering
   \includegraphics[width=0.9\linewidth]{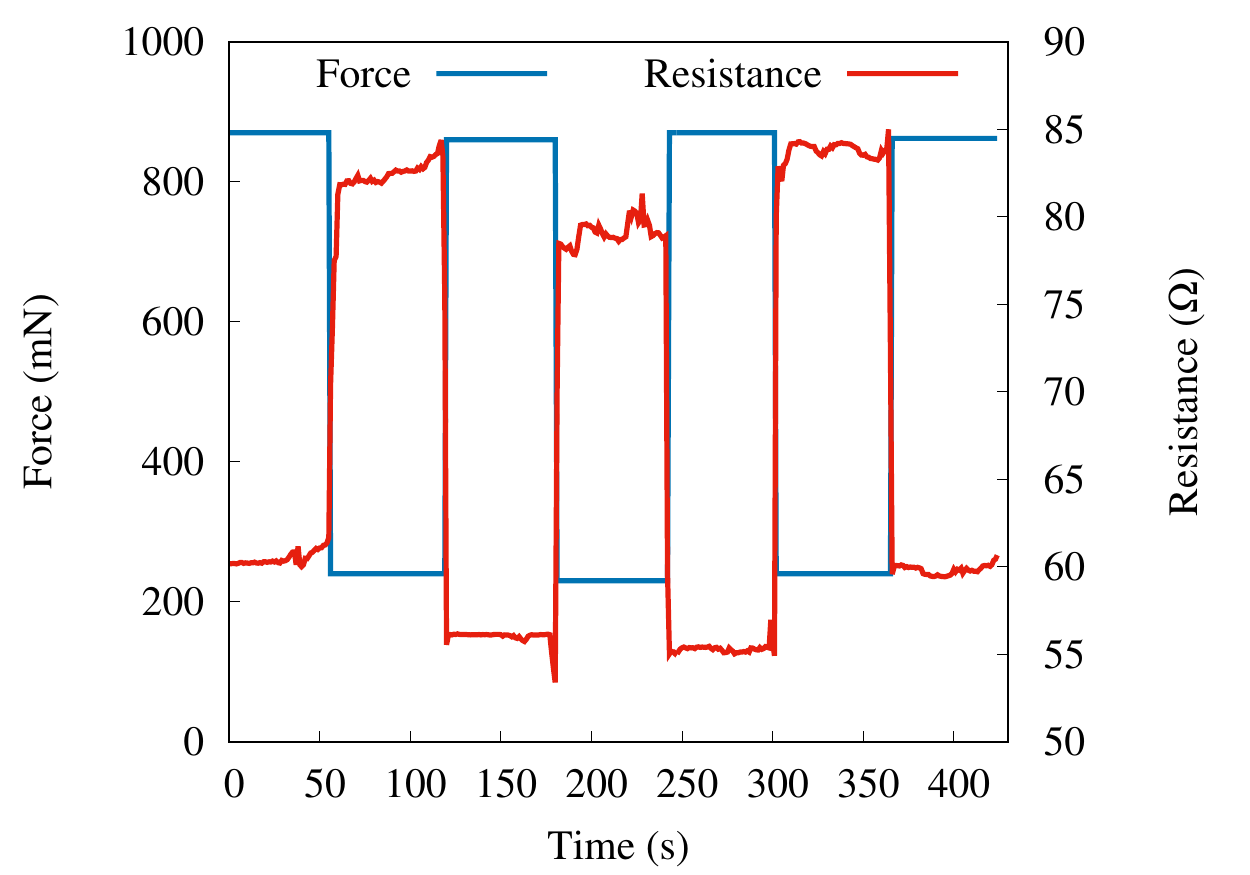}
    \caption{SiNWs PZR response due to repeated load-unload force in real time for sample D. }
    \label{fig:load}
\end{figure}

Up to \SI{35}{\%} change in resistivity was observed over the pressure variation in the range of \SIrange{0.1}{0.9}{\newton}.
In order to test the stability of SiNWs PZR response, we perform several sets of measurements over different samples.
In this stability test the resistance shift was measured in periods of \SI{60}{\second} of loading and unloading 
forcet, for all the samples.
All samples were loaded by \SI{860+-10}{\milli\newton} and then unloaded to \SI{220+-10}{\milli\newton}. We observe fast 
response, good stability and high sensitivity to repeated pressure changes. The results of a 
reproducibility test for sample D is presented in \autoref{fig:load}.

\begin{figure}
    \centering
    \includegraphics[width=0.9\linewidth]{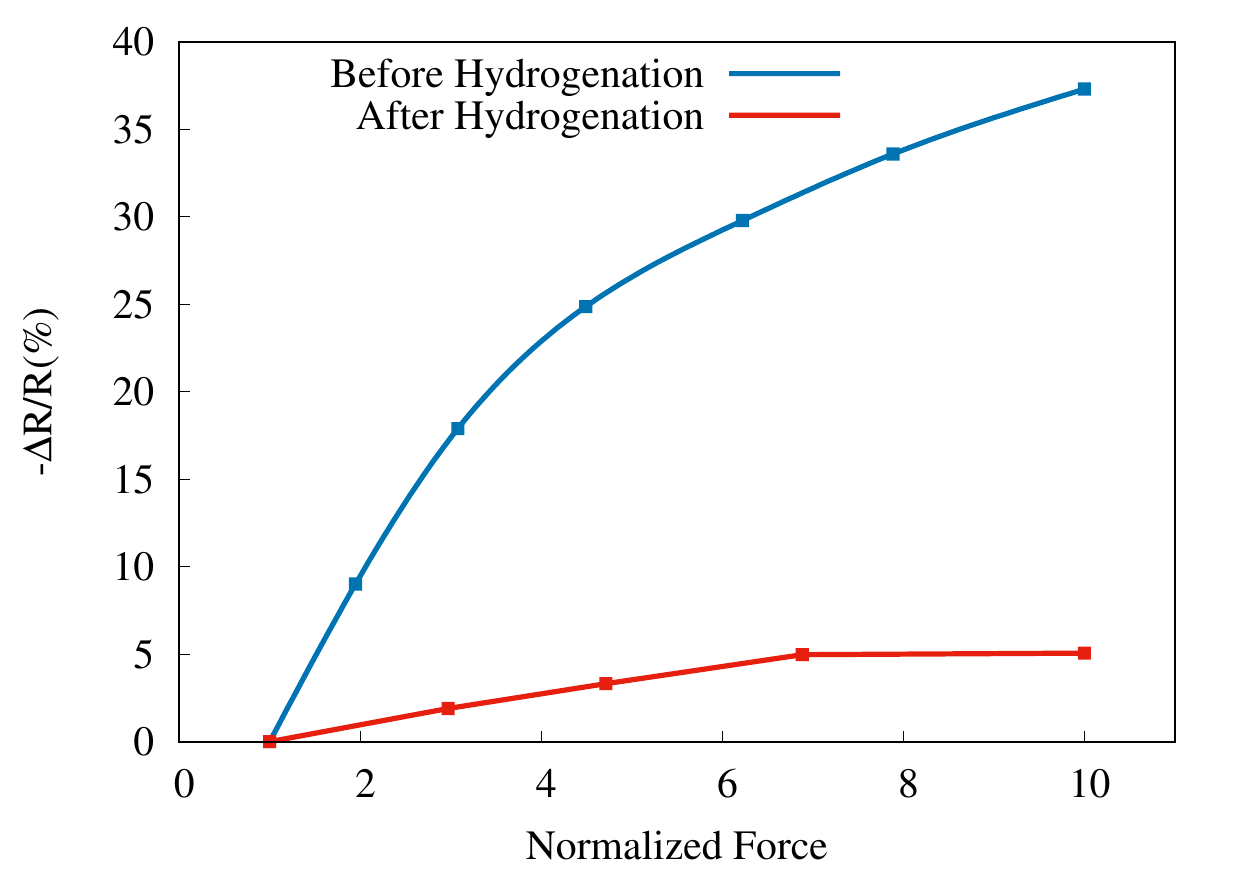}
    \caption{SiNWs PZR response for sample D before and after hydrogenation. }
    \label{fig:hydrogen}
\end{figure}

Hydrogenation (exposure to hydrogen plasma) is widely used in the electronic industry to increase the mobility of charge carriers in semiconductors.
It neutralizes deep and shallow defects and charged surface states \cite{danielsson2010effect,svavarsson2020hydrogenated}.
Because the PZR effect has been attributed to surface states, we applied a hydrogenation treatment in order to explore 
the origin of the PZR effect in SiNWs.  
For hydrogen plasma treatment we used a custom built inductively coupled discharge setup with cylindrical geometry (\SI{290}{\milli\meter} long quartz tube with diameter of \SI{34}{\milli\meter}).
The quartz tube was held inside a circular copper inductive coil with diameter of \SI{54}{\milli\meter}.
A radio frequency power generator CERSAR (c) (\SI{13.56}{\mega\hertz}) source coupled with impedance matching unit was utilized.
For hydrogenation a gas mixture of Ar/H\textsubscript{2} (\SI{30/70}{\%}) was used and the throttle valves were adjusted to stabilize the gas pressure of \SI{29}{\milli\bar}.
A more detailed description and schematic of the hydrogenation setup can be found elsewhere \cite{sultan2019enhanced,sultan2018effect}.
In \autoref{fig:hydrogen} we compare the behavior of sample D before and after the hydrogenation.
The PZR effect decreases dramatically, and can be attributed to passivation of the surface states.

\subsection{Electrical response under isostatic pressure}
\begin{figure}
    \centering
 \includegraphics[width=1\linewidth]{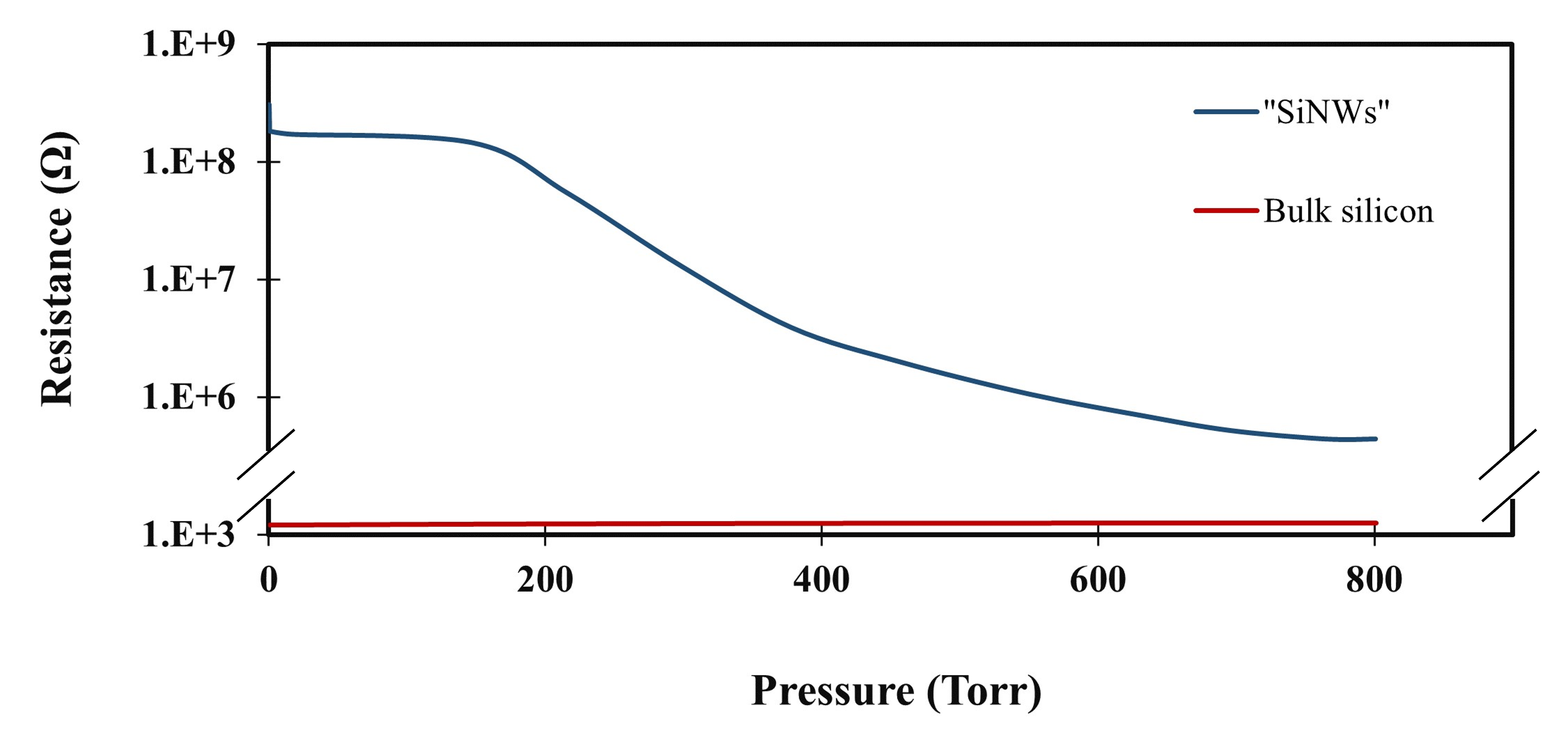}
    \caption{The resistance of sample E (on a log 10 scale) as a function of pressure measured in the vacuum chamber, is shown with the blue curve. For comparison, in red color, the resistance of a bulk silicon sample (i.\ e. without nanowires)}
    \label{fig:r_vs_p}
\end{figure}

Sample E (\SI{10}{\micro\meter}) was used for the investigation of the PZR response to isostatic pressure in 
a vacuum chamber.
The PZR response was tested in the chamber under re-pressurizing condition in the range of \SIrange{e-4}{e3}{\milli\bar}.
The results are shown in \autoref{fig:r_vs_p} where the resistance is plotted as a function of the pressure measured at a fixed bias voltage of \SI{5}{\volt}.
We observed a dramatic increase of the resistance, by more than two orders of magnitude, when pumping out the air from the 
vacuum chamber. We believe this is a result of combined mechanical and chemical effects: the pressure drop removes 
mechanical stress, and the lack of air and humidity suppresses the chemisorption.  Note that the resistance values for sample E are much higher than for the samples A-D because of the co-planar configuration of the contacts and also 
because of the larger length of the nanowires. 

In the Appendix we show, for comparison, the behavior of periodic arrays of SiNWs with increasing isostatic pressure.  

\section{Discussion}

For solid materials the inter-atomic spacing may be altered by strain.
Consequently, apart from geometric changes, in semiconductors, bandstructure related details, such as bandgap, or effective mass, may change and thereby the resistivity as well.
Within a certain range of strain this relationship is linear \cite{Matsuda1993,Rowe2014}.
When a uniaxial stress $X$ is applied, the piezoresistance coefficient of the resistivity $\rho$ in the direction of stress, is defined as  
\begin{equation}
\pi_l=\frac{\Delta\rho}{\rho_0}\frac{1}{X}   \ ,
\label{eq:x1}
\end{equation}
where $\Delta \rho$ is the stress-induced change in the resistivity and $\rho_0$ is the reference resistivity of the unstressed material. 

In our case a uniaxial compressive force $F$ was applied on SiNWs along their length by a force meter.
The stress is $X=F/A_{\mathrm t}$, $A_{\mathrm t}$ being the total cross sectional area of the nanowires, which is equal to $pA_{\mathrm m}$, where $A_{\mathrm m}$ is the cross-sectional area of the pin pressed into the wires, and $p$ is the relative cross-sectional area of the wires.
Assuming the electrical resistivity of the nanowires is proportional to their resistance \autoref{eq:x1} becomes

\begin{equation}
   \pi_{l}= \frac{p A_{\mathrm m}}{F}  \times \frac{\Delta R}{R_0}\ .
    \label{eq:our_final_formula}
\end{equation}

The structure of our SiNWs array is robust and stable as the wires are partially interconnected, which provides high structural strength.
Such stability and fast response of SiNWs is a desired  property for many devices such as solid state accelerometers and bipolar transistors \cite{Doll2013}.
Also our results are in agreement with the study of Ghosh et al.~\cite{Ghosh2021}, in which large diameter Si nanorod-based sensors was used for force detection.

He et al. \cite{He2006}, who measured PZR coefficients of single (p-doped) SiNWs, with diameters \SIrange{50}{300}{\nano\meter}, made from wafers with resistivities between \SIrange{0.003}{10}{\ohm\cm}, found that the PZR was roughly inversely proportional to the diameter and proportional to the wafer resistivity.
For a similar diameter and wafer resistivity as in our present work, \SI{\sim 150}{\nm} and \SI{10}{\ohm\cm}, respectively, the PZR coefficient $\pi_l$ for a single wire was of the order of \SI{e-7}{\per\pascal}.
(According to Figure 2d of Ref. \onlinecite{He2006}.) 
Using \autoref{eq:our_final_formula}, with $p$\num{= 0.28}, $A_m$\SI{= 49}{\mm\squared}, $F$\SI{= 0.8}{\newton}, and $\Delta R /R$\num{= 0.35}, we obtain $\pi_l$\SI{\approx 6e-6}{\per\pascal}, i.\ e. almost two orders of magnitude higher.
Note that in principle \autoref{eq:x1} does not depend on sample details such as the number of nanowires or their configuration, which we believe plays a role in our case.
Therefore we attribute this higher value to a collective PZR effect brought about by the interaction between multiple and interconnected wires, rather than the response of a single nanowire.  
\section{Conclusions}
In summary, large arrays of interconnected SiNWs were fabricated in a simple tree step wet chemical process and 
used for testing piezoresistance effect in nanowires.
The interconnected structures of the SiNWs provide a great increase in mechanical stability.
A pressure change of \SI{100}{\pascal} could be measured with this robust device.
The calculated PZR coefficient based on SiNWs array with NWs length of \SI{3}{\micro\meter} as resulted after MACE 
etched for \SI{7}{\minute}, sample D, was almost two orders of magnitude higher for our sensor than reported for a 
single SiNW.
Repeated measurements for different samples fabricated with the same process demonstrated good reproducibility with  
less than \si{5}{\%} deviation in pressure sensing.
The electrical resistance of SiNWs of \SI{10}{\micro\meter} length increased more than two orders of magnitude when 
measured in vacuum. 
This findings make the device based on random and interconnected SiNWs a strong candidate as a simple and inexpensive 
alternative to various pressure sensing applications.

\appendix*


\renewcommand{\thefigure}{A\arabic{figure}}
\setcounter{figure}{0}

\renewcommand{\theequation}{A\arabic{equation}}
\setcounter{equation}{0}

\section{Periodic nanowires}
\begin{figure}
  \centering
   \includegraphics[width=0.8\linewidth]{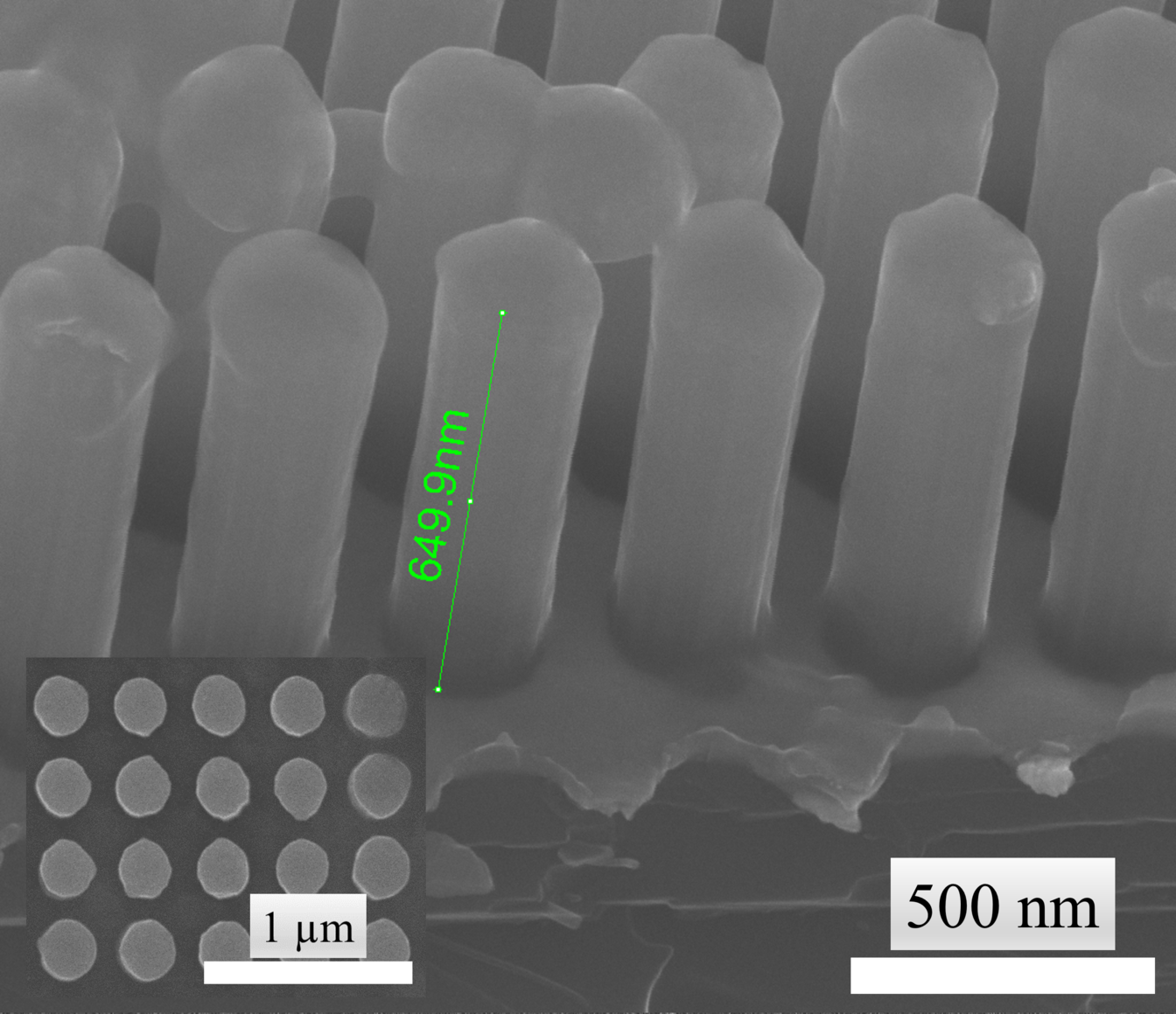}\\
  \caption{Cross-sectional SEM micrographs of periodic SiNWs-ZnO array with wire length of \SI{650}{\nm}, inset shows the top view image.}
   \label{fig:Periodic_SiNWs_SEM}
\end{figure}

In this Section we show the samples with periodic arrays of nanowires and the corresponding results, to be compared with the random wires presented in the main text.
These samples were processed in \SI{10}{\ohm\cm} p-type (100) Si substrates, by chemical etching method, using 1:1 H\textsubscript{2}0: HF \SI{4}{\milli\liter}+\SI{0.1}{\milli\liter} H\textsubscript{2}O\textsubscript{2} solution.
Vertical nanowires with length/diameter of \SI{650}{\nm}/\SI{272}{\nm}, as given by SEM investigations, resulted when the etching time was \SI{5}{\minute}.
In order to achieve the electrical contact to the SiNW arrays, Al\textsubscript{2}O\textsubscript{3} (\si{2}{\%}) doped ZnO films (about \SI{43}{\nm} thick and resistivity of about \SI{210}{\ohm\cm}) were deposited on the samples by RF magnetron sputtering using an Oxford PlasmaLab System 400 equipment.
The electrodes consist of copper wires connected to the SiNW arrays and also to Si substrate by silver conductive glue paste.
SEM images of the periodic SiNWs are shown in \autoref{fig:Periodic_SiNWs_SEM}.

\begin{figure}
  \centering
    \includegraphics[width=1\linewidth]{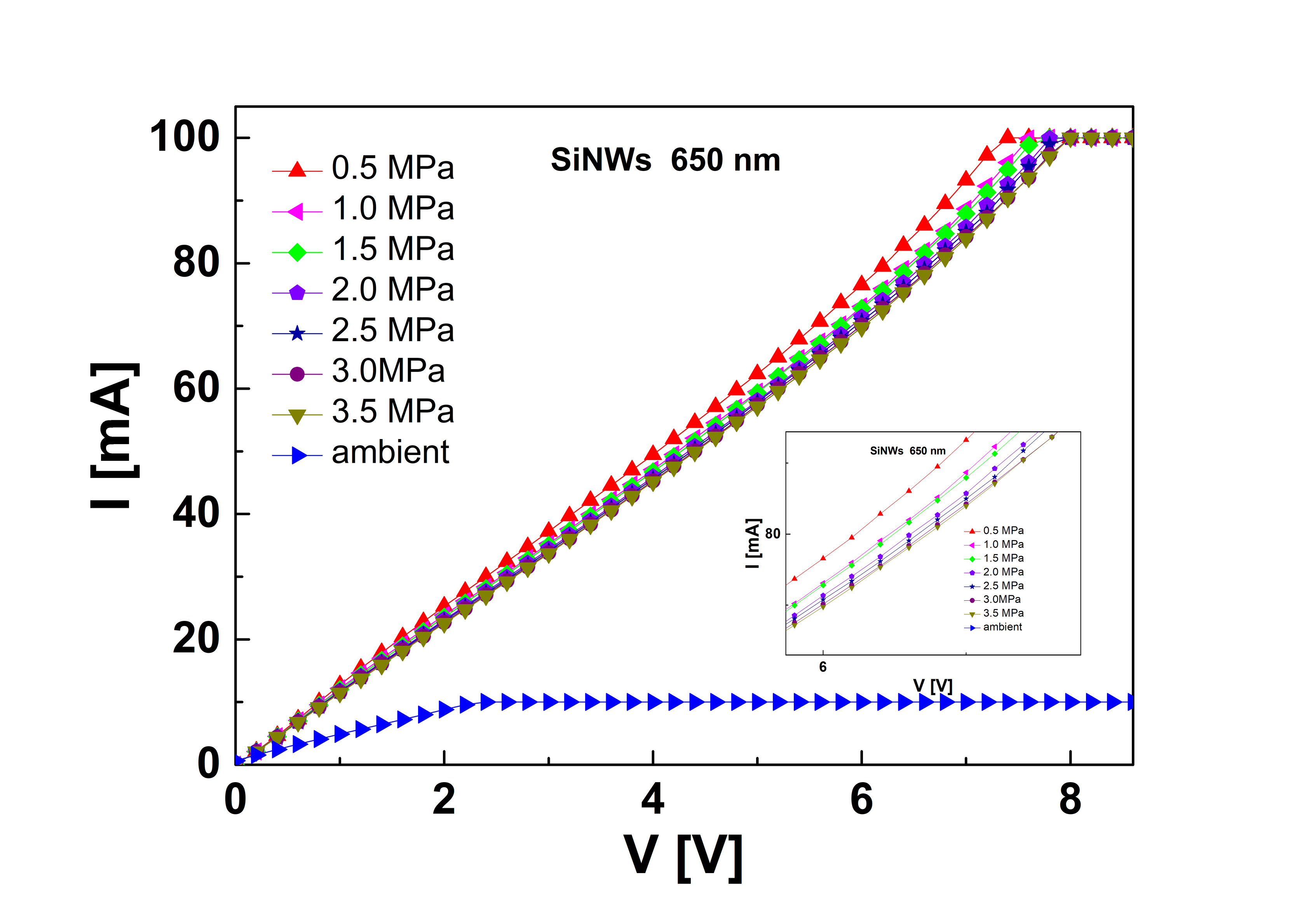}
    (a)
  \includegraphics[width=1\linewidth]{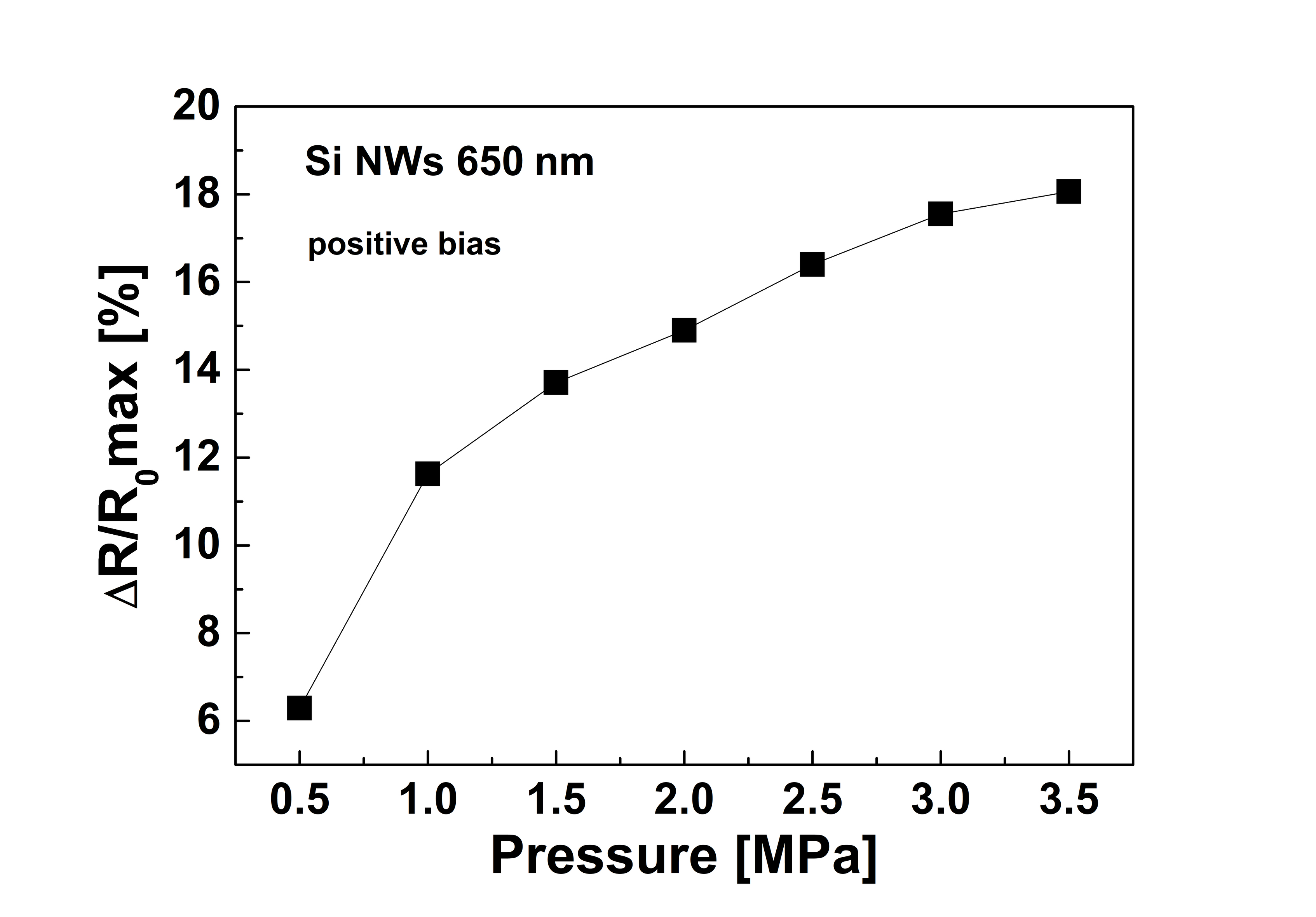}
  (b)
  \caption{(a) I-V characteristics for direct bias of the \SI{650}{\nm} periodic SiNWs, with applied pressure in the range \SIrange{0.5}{3.5}{\mega\pascal}. The blue curve corresponds to atmospheric pressure. (b) Variation of $\Delta R/R_0$ with applied pressure for a fixed direct bias of \SI{2}{\volt}. }
   \label{fig:Periodic_SiNWs_PZR}
\end{figure}

Isostatic pressure in the range \SIrange{1}{35}{\bar} (\SIrange{0.1}{3.5}{\mega\pascal}) was applied on this sample by placing it in a controlled pressure chamber Type 4E/2lt, Büchiglasuster- Switzerland, under nitrogen atmosphere.
The relative resistance variation $\Delta R/R_0$  was in the interval  \SIrange{6}{18}{\%} at pressures between \SIrange{0.5}{3.5}{\mega\pascal} at direct bias, as shown in \autoref{fig:Periodic_SiNWs_PZR}.
The PZR had only a slight dependence on applied pressure for reversed bias (not shown).

\clearpage


\acknowledgments{This work was supported by Reykjavik University Ph.D. fund No. 220006, and funding from the Icelandic Research Fund Grant No. 218029-051. RP acknowledges support from the Romanian Core Program Contract No.14 N/2019 Ministry of Research, Innovation, and Digitalization.}

\bibliographystyle{apsrev}
\bibliography{Bibliography}

\begin{thebibliography}{33}
\expandafter\ifx\csname natexlab\endcsname\relax\def\natexlab#1{#1}\fi
\expandafter\ifx\csname bibnamefont\endcsname\relax
  \def\bibnamefont#1{#1}\fi
\expandafter\ifx\csname bibfnamefont\endcsname\relax
  \def\bibfnamefont#1{#1}\fi
\expandafter\ifx\csname citenamefont\endcsname\relax
  \def\citenamefont#1{#1}\fi
\expandafter\ifx\csname url\endcsname\relax
  \def\url#1{\texttt{#1}}\fi
\expandafter\ifx\csname urlprefix\endcsname\relax\def\urlprefix{URL }\fi
\providecommand{\bibinfo}[2]{#2}
\providecommand{\eprint}[2][]{\url{#2}}

\bibitem[{\citenamefont{Peng et~al.}(2013)\citenamefont{Peng, Wang, Li, Hu, and
  Lee}}]{Peng2013}
\bibinfo{author}{\bibfnamefont{K.-Q.} \bibnamefont{Peng}},
  \bibinfo{author}{\bibfnamefont{X.}~\bibnamefont{Wang}},
  \bibinfo{author}{\bibfnamefont{L.}~\bibnamefont{Li}},
  \bibinfo{author}{\bibfnamefont{Y.}~\bibnamefont{Hu}}, \bibnamefont{and}
  \bibinfo{author}{\bibfnamefont{S.-T.} \bibnamefont{Lee}},
  \bibinfo{journal}{Nano Today} \textbf{\bibinfo{volume}{8}},
  \bibinfo{pages}{75} (\bibinfo{year}{2013}).

\bibitem[{\citenamefont{Heris et~al.}(2020)\citenamefont{Heris, Kateb,
  Erlingsson, and Manolescu}}]{Heris2020}
\bibinfo{author}{\bibfnamefont{H.~R.} \bibnamefont{Heris}},
  \bibinfo{author}{\bibfnamefont{M.}~\bibnamefont{Kateb}},
  \bibinfo{author}{\bibfnamefont{S.~I.} \bibnamefont{Erlingsson}},
  \bibnamefont{and}
  \bibinfo{author}{\bibfnamefont{A.}~\bibnamefont{Manolescu}},
  \bibinfo{journal}{Nanotechnology} \textbf{\bibinfo{volume}{31}},
  \bibinfo{pages}{424006} (\bibinfo{year}{2020}).

\bibitem[{\citenamefont{Heris et~al.}(2022)\citenamefont{Heris, Kateb,
  Erlingsson, and Manolescu}}]{Heris2022}
\bibinfo{author}{\bibfnamefont{H.~R.} \bibnamefont{Heris}},
  \bibinfo{author}{\bibfnamefont{M.}~\bibnamefont{Kateb}},
  \bibinfo{author}{\bibfnamefont{S.~I.} \bibnamefont{Erlingsson}},
  \bibnamefont{and}
  \bibinfo{author}{\bibfnamefont{A.}~\bibnamefont{Manolescu}},
  \bibinfo{journal}{Surfaces and Interfaces} p. \bibinfo{pages}{101834}
  (\bibinfo{year}{2022}).

\bibitem[{\citenamefont{Zhou et~al.}(2003)\citenamefont{Zhou, Hu, Li, Ma, Lee,
  and Lee}}]{Zhou2003}
\bibinfo{author}{\bibfnamefont{X.}~\bibnamefont{Zhou}},
  \bibinfo{author}{\bibfnamefont{J.}~\bibnamefont{Hu}},
  \bibinfo{author}{\bibfnamefont{C.}~\bibnamefont{Li}},
  \bibinfo{author}{\bibfnamefont{D.}~\bibnamefont{Ma}},
  \bibinfo{author}{\bibfnamefont{C.}~\bibnamefont{Lee}}, \bibnamefont{and}
  \bibinfo{author}{\bibfnamefont{S.}~\bibnamefont{Lee}},
  \bibinfo{journal}{Chemical Physics Letters} \textbf{\bibinfo{volume}{369}},
  \bibinfo{pages}{220} (\bibinfo{year}{2003}).

\bibitem[{\citenamefont{Peng et~al.}(2009)\citenamefont{Peng, Wang, and
  Lee}}]{Peng2009}
\bibinfo{author}{\bibfnamefont{K.-Q.} \bibnamefont{Peng}},
  \bibinfo{author}{\bibfnamefont{X.}~\bibnamefont{Wang}}, \bibnamefont{and}
  \bibinfo{author}{\bibfnamefont{S.-T.} \bibnamefont{Lee}},
  \bibinfo{journal}{Applied Physics Letters} \textbf{\bibinfo{volume}{95}},
  \bibinfo{pages}{243112} (\bibinfo{year}{2009}).

\bibitem[{\citenamefont{Smith}(1954)}]{Smith1954}
\bibinfo{author}{\bibfnamefont{C.~S.} \bibnamefont{Smith}},
  \bibinfo{journal}{Physics Review} \textbf{\bibinfo{volume}{94}},
  \bibinfo{pages}{42} (\bibinfo{year}{1954}).

\bibitem[{\citenamefont{Tufte et~al.}(1962)\citenamefont{Tufte, Chapman, and
  Long}}]{Tufte1962}
\bibinfo{author}{\bibfnamefont{O.~N.} \bibnamefont{Tufte}},
  \bibinfo{author}{\bibfnamefont{P.~D.} \bibnamefont{Chapman}},
  \bibnamefont{and} \bibinfo{author}{\bibfnamefont{D.}~\bibnamefont{Long}},
  \bibinfo{journal}{Journal of Applied Physics} \textbf{\bibinfo{volume}{33}}
  (\bibinfo{year}{1962}).

\bibitem[{\citenamefont{Tortonese et~al.}(1993)\citenamefont{Tortonese,
  Barrett, and Quate}}]{Tortonese1993}
\bibinfo{author}{\bibfnamefont{M.}~\bibnamefont{Tortonese}},
  \bibinfo{author}{\bibfnamefont{R.~C.} \bibnamefont{Barrett}},
  \bibnamefont{and} \bibinfo{author}{\bibfnamefont{C.~F.} \bibnamefont{Quate}},
  \bibinfo{journal}{Applied Physics Letters} \textbf{\bibinfo{volume}{62}},
  \bibinfo{pages}{834} (\bibinfo{year}{1993}).

\bibitem[{\citenamefont{Ning et~al.}(1995)\citenamefont{Ning, Loke, and
  McKinnon}}]{Ning1995}
\bibinfo{author}{\bibfnamefont{Y.}~\bibnamefont{Ning}},
  \bibinfo{author}{\bibfnamefont{Y.}~\bibnamefont{Loke}}, \bibnamefont{and}
  \bibinfo{author}{\bibfnamefont{G.}~\bibnamefont{McKinnon}},
  \bibinfo{journal}{Sensors and Actuators A: Physical}
  \textbf{\bibinfo{volume}{48}}, \bibinfo{pages}{55} (\bibinfo{year}{1995}).

\bibitem[{\citenamefont{Wee et~al.}(2005)\citenamefont{Wee, Kang, Park, Kang,
  Yoon, Park, and Kim}}]{Wee2005}
\bibinfo{author}{\bibfnamefont{K.}~\bibnamefont{Wee}},
  \bibinfo{author}{\bibfnamefont{G.}~\bibnamefont{Kang}},
  \bibinfo{author}{\bibfnamefont{J.}~\bibnamefont{Park}},
  \bibinfo{author}{\bibfnamefont{J.}~\bibnamefont{Kang}},
  \bibinfo{author}{\bibfnamefont{D.}~\bibnamefont{Yoon}},
  \bibinfo{author}{\bibfnamefont{J.}~\bibnamefont{Park}}, \bibnamefont{and}
  \bibinfo{author}{\bibfnamefont{T.}~\bibnamefont{Kim}},
  \bibinfo{journal}{Biosensors and Bioelectronics}
  \textbf{\bibinfo{volume}{20}}, \bibinfo{pages}{1932} (\bibinfo{year}{2005}).

\bibitem[{\citenamefont{Tiwari et~al.}(2021)\citenamefont{Tiwari, Billot,
  Cl{\'e}vy, Agnus, Piat, and Lutz}}]{Tiwari2021}
\bibinfo{author}{\bibfnamefont{B.}~\bibnamefont{Tiwari}},
  \bibinfo{author}{\bibfnamefont{M.}~\bibnamefont{Billot}},
  \bibinfo{author}{\bibfnamefont{C.}~\bibnamefont{Cl{\'e}vy}},
  \bibinfo{author}{\bibfnamefont{J.}~\bibnamefont{Agnus}},
  \bibinfo{author}{\bibfnamefont{E.}~\bibnamefont{Piat}}, \bibnamefont{and}
  \bibinfo{author}{\bibfnamefont{P.}~\bibnamefont{Lutz}},
  \bibinfo{journal}{Sensors} \textbf{\bibinfo{volume}{21}},
  \bibinfo{pages}{6059} (\bibinfo{year}{2021}).

\bibitem[{\citenamefont{He and Yang}(2006)}]{He2006}
\bibinfo{author}{\bibfnamefont{R.}~\bibnamefont{He}} \bibnamefont{and}
  \bibinfo{author}{\bibfnamefont{P.}~\bibnamefont{Yang}},
  \bibinfo{journal}{Nature nanotechnology} \textbf{\bibinfo{volume}{1}},
  \bibinfo{pages}{42} (\bibinfo{year}{2006}).

\bibitem[{\citenamefont{Gao et~al.}(2017)\citenamefont{Gao, Yang, Zheng, and
  Kun}}]{Gao2017}
\bibinfo{author}{\bibfnamefont{D.}~\bibnamefont{Gao}},
  \bibinfo{author}{\bibfnamefont{Z.}~\bibnamefont{Yang}},
  \bibinfo{author}{\bibfnamefont{L.}~\bibnamefont{Zheng}}, \bibnamefont{and}
  \bibinfo{author}{\bibfnamefont{Z.}~\bibnamefont{Kun}},
  \bibinfo{journal}{Nanotechnology} \textbf{\bibinfo{volume}{28}},
  \bibinfo{pages}{095702} (\bibinfo{year}{2017}).

\bibitem[{\citenamefont{Zhang et~al.}(2012)\citenamefont{Zhang, Lou, and
  Lee}}]{Zhang2012}
\bibinfo{author}{\bibfnamefont{S.}~\bibnamefont{Zhang}},
  \bibinfo{author}{\bibfnamefont{L.}~\bibnamefont{Lou}}, \bibnamefont{and}
  \bibinfo{author}{\bibfnamefont{C.}~\bibnamefont{Lee}},
  \bibinfo{journal}{Applied Physics Letters} \textbf{\bibinfo{volume}{100}},
  \bibinfo{pages}{023111} (\bibinfo{year}{2012}).

\bibitem[{\citenamefont{Cheng et~al.}(2018)\citenamefont{Cheng, Yu, Kong, Yu,
  Wang, Ma, Wang, Wang, Pan, and Shi}}]{Cheng2018}
\bibinfo{author}{\bibfnamefont{W.}~\bibnamefont{Cheng}},
  \bibinfo{author}{\bibfnamefont{L.}~\bibnamefont{Yu}},
  \bibinfo{author}{\bibfnamefont{D.}~\bibnamefont{Kong}},
  \bibinfo{author}{\bibfnamefont{Z.}~\bibnamefont{Yu}},
  \bibinfo{author}{\bibfnamefont{H.}~\bibnamefont{Wang}},
  \bibinfo{author}{\bibfnamefont{Z.}~\bibnamefont{Ma}},
  \bibinfo{author}{\bibfnamefont{Y.}~\bibnamefont{Wang}},
  \bibinfo{author}{\bibfnamefont{J.}~\bibnamefont{Wang}},
  \bibinfo{author}{\bibfnamefont{L.}~\bibnamefont{Pan}}, \bibnamefont{and}
  \bibinfo{author}{\bibfnamefont{Y.}~\bibnamefont{Shi}}, \bibinfo{journal}{IEEE
  Electron Device Letters} \textbf{\bibinfo{volume}{39}}, \bibinfo{pages}{1069}
  (\bibinfo{year}{2018}).

\bibitem[{\citenamefont{Nguyen and Lee}(2021)}]{Nguyen2021}
\bibinfo{author}{\bibfnamefont{T.-D.} \bibnamefont{Nguyen}} \bibnamefont{and}
  \bibinfo{author}{\bibfnamefont{J.~S.} \bibnamefont{Lee}},
  \bibinfo{journal}{Sensors} \textbf{\bibinfo{volume}{22}}, \bibinfo{pages}{50}
  (\bibinfo{year}{2021}).

\bibitem[{\citenamefont{Kim et~al.}(2020)\citenamefont{Kim, Ahn, and
  Ji}}]{Kim2020}
\bibinfo{author}{\bibfnamefont{C.}~\bibnamefont{Kim}},
  \bibinfo{author}{\bibfnamefont{H.}~\bibnamefont{Ahn}}, \bibnamefont{and}
  \bibinfo{author}{\bibfnamefont{T.}~\bibnamefont{Ji}}, \bibinfo{journal}{IEEE
  Electron Device Letters} \textbf{\bibinfo{volume}{41}}, \bibinfo{pages}{1233}
  (\bibinfo{year}{2020}).

\bibitem[{\citenamefont{Ghosh et~al.}(2021)\citenamefont{Ghosh, Song, Park,
  Tchoe, Guha, Lee, Lim, Kim, Kim, Kim et~al.}}]{Ghosh2021}
\bibinfo{author}{\bibfnamefont{R.}~\bibnamefont{Ghosh}},
  \bibinfo{author}{\bibfnamefont{M.~S.} \bibnamefont{Song}},
  \bibinfo{author}{\bibfnamefont{J.}~\bibnamefont{Park}},
  \bibinfo{author}{\bibfnamefont{Y.}~\bibnamefont{Tchoe}},
  \bibinfo{author}{\bibfnamefont{P.}~\bibnamefont{Guha}},
  \bibinfo{author}{\bibfnamefont{W.}~\bibnamefont{Lee}},
  \bibinfo{author}{\bibfnamefont{Y.}~\bibnamefont{Lim}},
  \bibinfo{author}{\bibfnamefont{B.}~\bibnamefont{Kim}},
  \bibinfo{author}{\bibfnamefont{S.-W.} \bibnamefont{Kim}},
  \bibinfo{author}{\bibfnamefont{M.}~\bibnamefont{Kim}}, \bibnamefont{et~al.},
  \bibinfo{journal}{Nano Energy} \textbf{\bibinfo{volume}{80}},
  \bibinfo{pages}{105537} (\bibinfo{year}{2021}).

\bibitem[{\citenamefont{Shiri et~al.}(2008)\citenamefont{Shiri, Kong, Buin, and
  Anantram}}]{shiri2008}
\bibinfo{author}{\bibfnamefont{D.}~\bibnamefont{Shiri}},
  \bibinfo{author}{\bibfnamefont{Y.}~\bibnamefont{Kong}},
  \bibinfo{author}{\bibfnamefont{A.}~\bibnamefont{Buin}}, \bibnamefont{and}
  \bibinfo{author}{\bibfnamefont{M.~P.~.} \bibnamefont{Anantram}},
  \bibinfo{journal}{Applied Physics Letters} \textbf{\bibinfo{volume}{93}},
  \bibinfo{pages}{07314} (\bibinfo{year}{2008}).

\bibitem[{\citenamefont{Zhang et~al.}(2016)\citenamefont{Zhang, Zhao, Ge, Li,
  Yang, and Mao}}]{Zhang2016}
\bibinfo{author}{\bibfnamefont{J.}~\bibnamefont{Zhang}},
  \bibinfo{author}{\bibfnamefont{Y.}~\bibnamefont{Zhao}},
  \bibinfo{author}{\bibfnamefont{Y.}~\bibnamefont{Ge}},
  \bibinfo{author}{\bibfnamefont{M.}~\bibnamefont{Li}},
  \bibinfo{author}{\bibfnamefont{L.}~\bibnamefont{Yang}}, \bibnamefont{and}
  \bibinfo{author}{\bibfnamefont{X.}~\bibnamefont{Mao}},
  \bibinfo{journal}{Micromachines} \textbf{\bibinfo{volume}{7}},
  \bibinfo{pages}{187} (\bibinfo{year}{2016}).

\bibitem[{\citenamefont{Rowe}(2014)}]{Rowe2014}
\bibinfo{author}{\bibfnamefont{A.}~\bibnamefont{Rowe}},
  \bibinfo{journal}{Journal of Materials Research}
  \textbf{\bibinfo{volume}{29}}, \bibinfo{pages}{731} (\bibinfo{year}{2014}).

\bibitem[{\citenamefont{Toriyama et~al.}(2002)\citenamefont{Toriyama, Tanimoto,
  and Sugiyama}}]{Toriyama2002}
\bibinfo{author}{\bibfnamefont{T.}~\bibnamefont{Toriyama}},
  \bibinfo{author}{\bibfnamefont{Y.}~\bibnamefont{Tanimoto}}, \bibnamefont{and}
  \bibinfo{author}{\bibfnamefont{S.}~\bibnamefont{Sugiyama}},
  \bibinfo{journal}{Journal of microelectromechanical systems}
  \textbf{\bibinfo{volume}{11}}, \bibinfo{pages}{605} (\bibinfo{year}{2002}).

\bibitem[{\citenamefont{Toriyama and Sugiyama}(2003)}]{Toriyama2003}
\bibinfo{author}{\bibfnamefont{T.}~\bibnamefont{Toriyama}} \bibnamefont{and}
  \bibinfo{author}{\bibfnamefont{S.}~\bibnamefont{Sugiyama}},
  \bibinfo{journal}{Sensors and Actuators A: Physical}
  \textbf{\bibinfo{volume}{108}}, \bibinfo{pages}{244} (\bibinfo{year}{2003}).

\bibitem[{\citenamefont{Schmidt et~al.}(2009)\citenamefont{Schmidt, Wittemann,
  Senz, and G{\"o}sele}}]{Schmidt2009}
\bibinfo{author}{\bibfnamefont{V.}~\bibnamefont{Schmidt}},
  \bibinfo{author}{\bibfnamefont{J.~V.} \bibnamefont{Wittemann}},
  \bibinfo{author}{\bibfnamefont{S.}~\bibnamefont{Senz}}, \bibnamefont{and}
  \bibinfo{author}{\bibfnamefont{U.}~\bibnamefont{G{\"o}sele}},
  \bibinfo{journal}{Advanced Materials} \textbf{\bibinfo{volume}{21}},
  \bibinfo{pages}{2681} (\bibinfo{year}{2009}).

\bibitem[{\citenamefont{Svavarsson~H et~al.}(2016)\citenamefont{Svavarsson~H,
  Hallgrimsson, Niraula, Lee~K, and Magnusson}}]{svavarsson2016}
\bibinfo{author}{\bibfnamefont{G.}~\bibnamefont{Svavarsson~H}},
  \bibinfo{author}{\bibfnamefont{H.}~\bibnamefont{Hallgrimsson},
  \bibfnamefont{B}}, \bibinfo{author}{\bibfnamefont{M.}~\bibnamefont{Niraula}},
  \bibinfo{author}{\bibfnamefont{J.}~\bibnamefont{Lee~K}}, \bibnamefont{and}
  \bibinfo{author}{\bibfnamefont{R.}~\bibnamefont{Magnusson}},
  \bibinfo{journal}{Applied Physics A-Materials Science and Processing}
  \textbf{\bibinfo{volume}{122}} (\bibinfo{year}{2016}).

\bibitem[{\citenamefont{Fakhri et~al.}(2021)\citenamefont{Fakhri, Sultan,
  Manolescu, Ingvarsson, Plugaru, Plugaru, and Svavarsson}}]{Fakhri2021}
\bibinfo{author}{\bibfnamefont{E.}~\bibnamefont{Fakhri}},
  \bibinfo{author}{\bibfnamefont{M.}~\bibnamefont{Sultan}},
  \bibinfo{author}{\bibfnamefont{A.}~\bibnamefont{Manolescu}},
  \bibinfo{author}{\bibfnamefont{S.}~\bibnamefont{Ingvarsson}},
  \bibinfo{author}{\bibfnamefont{N.}~\bibnamefont{Plugaru}},
  \bibinfo{author}{\bibfnamefont{R.}~\bibnamefont{Plugaru}}, \bibnamefont{and}
  \bibinfo{author}{\bibfnamefont{H.}~\bibnamefont{Svavarsson}}, in
  \emph{\bibinfo{booktitle}{2021 International Semiconductor Conference (CAS)}}
  (\bibinfo{organization}{IEEE}, \bibinfo{year}{2021}), pp.
  \bibinfo{pages}{147--150}.

\bibitem[{\citenamefont{Plugaru et~al.}(2022)\citenamefont{Plugaru, Fakhri,
  Romanitan, Mihalache, Craciun, Plugaru, \'Arnason, Sultan, Nemnes, Ingvarsson
  et~al.}}]{plugaru2022structure}
\bibinfo{author}{\bibfnamefont{R.}~\bibnamefont{Plugaru}},
  \bibinfo{author}{\bibfnamefont{E.}~\bibnamefont{Fakhri}},
  \bibinfo{author}{\bibfnamefont{C.}~\bibnamefont{Romanitan}},
  \bibinfo{author}{\bibfnamefont{I.}~\bibnamefont{Mihalache}},
  \bibinfo{author}{\bibfnamefont{G.}~\bibnamefont{Craciun}},
  \bibinfo{author}{\bibfnamefont{N.}~\bibnamefont{Plugaru}},
  \bibinfo{author}{\bibfnamefont{H.~O.} \bibnamefont{\'Arnason}},
  \bibinfo{author}{\bibfnamefont{M.~T.} \bibnamefont{Sultan}},
  \bibinfo{author}{\bibfnamefont{G.~A.} \bibnamefont{Nemnes}},
  \bibinfo{author}{\bibfnamefont{S.}~\bibnamefont{Ingvarsson}},
  \bibnamefont{et~al.} (\bibinfo{year}{2022}),
  \eprint{https://arxiv.org/abs/2206.05006}.

\bibitem[{\citenamefont{Danielsson et~al.}(2010)\citenamefont{Danielsson,
  Gudmundsson, and Svavarsson}}]{danielsson2010effect}
\bibinfo{author}{\bibfnamefont{D.}~\bibnamefont{Danielsson}},
  \bibinfo{author}{\bibfnamefont{J.}~\bibnamefont{Gudmundsson}},
  \bibnamefont{and}
  \bibinfo{author}{\bibfnamefont{H.}~\bibnamefont{Svavarsson}},
  \bibinfo{journal}{Physica Scripta} \textbf{\bibinfo{volume}{2010}},
  \bibinfo{pages}{014005} (\bibinfo{year}{2010}).

\bibitem[{\citenamefont{Svavarsson et~al.}(2020)\citenamefont{Svavarsson,
  Sultan, Lee, and Magnusson}}]{svavarsson2020hydrogenated}
\bibinfo{author}{\bibfnamefont{H.~G.} \bibnamefont{Svavarsson}},
  \bibinfo{author}{\bibfnamefont{M.~T.} \bibnamefont{Sultan}},
  \bibinfo{author}{\bibfnamefont{K.~J.} \bibnamefont{Lee}}, \bibnamefont{and}
  \bibinfo{author}{\bibfnamefont{R.}~\bibnamefont{Magnusson}}, in
  \emph{\bibinfo{booktitle}{2020 IEEE Research and Applications of Photonics}}
  (\bibinfo{organization}{IEEE}, \bibinfo{year}{2020}), pp.
  \bibinfo{pages}{1--2}.

\bibitem[{\citenamefont{Sultan et~al.}(2019)\citenamefont{Sultan, Gudmundsson,
  Manolescu, Stoica, Ciurea, and Svavarsson}}]{sultan2019enhanced}
\bibinfo{author}{\bibfnamefont{M.}~\bibnamefont{Sultan}},
  \bibinfo{author}{\bibfnamefont{J.~T.} \bibnamefont{Gudmundsson}},
  \bibinfo{author}{\bibfnamefont{A.}~\bibnamefont{Manolescu}},
  \bibinfo{author}{\bibfnamefont{T.}~\bibnamefont{Stoica}},
  \bibinfo{author}{\bibfnamefont{M.}~\bibnamefont{Ciurea}}, \bibnamefont{and}
  \bibinfo{author}{\bibfnamefont{H.}~\bibnamefont{Svavarsson}},
  \bibinfo{journal}{Applied Surface Science} \textbf{\bibinfo{volume}{479}},
  \bibinfo{pages}{403} (\bibinfo{year}{2019}).

\bibitem[{\citenamefont{Sultan et~al.}(2018)\citenamefont{Sultan, Gudmundsson,
  Manolescu, Ciurea, and Svavarsson}}]{sultan2018effect}
\bibinfo{author}{\bibfnamefont{M.}~\bibnamefont{Sultan}},
  \bibinfo{author}{\bibfnamefont{J.~T.} \bibnamefont{Gudmundsson}},
  \bibinfo{author}{\bibfnamefont{A.}~\bibnamefont{Manolescu}},
  \bibinfo{author}{\bibfnamefont{M.}~\bibnamefont{Ciurea}}, \bibnamefont{and}
  \bibinfo{author}{\bibfnamefont{H.}~\bibnamefont{Svavarsson}}, in
  \emph{\bibinfo{booktitle}{2018 International Semiconductor Conference (CAS)}}
  (\bibinfo{organization}{IEEE}, \bibinfo{year}{2018}), pp.
  \bibinfo{pages}{257--260}.

\bibitem[{\citenamefont{Matsuda et~al.}(1993)\citenamefont{Matsuda, Suzuki,
  Yamamura, and Kanda}}]{Matsuda1993}
\bibinfo{author}{\bibfnamefont{K.}~\bibnamefont{Matsuda}},
  \bibinfo{author}{\bibfnamefont{K.}~\bibnamefont{Suzuki}},
  \bibinfo{author}{\bibfnamefont{K.}~\bibnamefont{Yamamura}}, \bibnamefont{and}
  \bibinfo{author}{\bibfnamefont{Y.}~\bibnamefont{Kanda}},
  \bibinfo{journal}{Journal of applied physics} \textbf{\bibinfo{volume}{73}},
  \bibinfo{pages}{1838} (\bibinfo{year}{1993}).

\bibitem[{\citenamefont{Doll and Pruitt}(2013)}]{Doll2013}
\bibinfo{author}{\bibfnamefont{J.~C.} \bibnamefont{Doll}} \bibnamefont{and}
  \bibinfo{author}{\bibfnamefont{B.~L.} \bibnamefont{Pruitt}},
  \emph{\bibinfo{title}{Piezoresistor design and applications}}
  (\bibinfo{publisher}{Springer}, \bibinfo{year}{2013}).

\end{thebibliography}

\end{document}